\documentclass[12pt,preprint]{aastex}
\usepackage{natbib}
\usepackage{graphicx,rotating}
\usepackage{amssymb}
\usepackage{epstopdf}
\usepackage{amsmath}
\usepackage{natbib}
\usepackage{lscape}
\usepackage{adjustbox,lipsum}
\newcommand{\beq}{\begin{equation}}
\newcommand{\eeq}{\end{equation}}
\def\etal{{\sl et~al.~}}

\newcommand{\HST}{{\it HST}}
\newcommand{\HSTs}{{\it HST~}}
\newcommand{\HIP}{{\it Hipparcos}}
\newcommand{\G}{{\it Gaia}}

\newcommand{\msini}{$\cal{M}$\,sin\,{i}~}

\newcommand{\m}{$\cal{M}$}
\newcommand{\mjup}{$\cal{M}_{\rm Jup}~$}
\newcommand{\mjupe}{$\cal{M}_{\rm Jup}$}

\newcommand{\msune}{$\cal{M}_{\odot}$}

\received{}
\revised{}
\accepted{}

\shorttitle{FGS Astrometry}
\shortauthors{Benedict et al.}

\begin{document}
\bibliographystyle{/Active/my2}

\title{Astrometry with {\it Hubble Space Telescope} Fine Guidance Sensors \\A Review\footnote{Based on observations made with the NASA/ESA Hubble Space Telescope, obtained at the Space Telescope Science Institute, which is operated by the Association of Universities for Research in Astronomy, Inc., under NASA contract NAS5-26555. } }

\author{ G.\ Fritz Benedict\altaffilmark{2}, Barbara E.
McArthur\altaffilmark{2}, Edmund P. Nelan\altaffilmark{3}, and Thomas E. Harrison\altaffilmark{4}}

\altaffiltext{2}{McDonald Observatory, University of Texas, Austin, TX 78712}

\altaffiltext{4}{Department of Astronomy, New Mexico State University, Box 30001, MSC 4500, Las Cruces, NM 88003-8001}
\altaffiltext{3}{Space Telescope Science Institute, 3700 San Martin Dr., Baltimore, MD 21218}




\begin{abstract}
Over the last 20 years {\it Hubble Space Telescope} Fine Guidance Sensor interferometric astrometry has produced precise and accurate parallaxes
of astrophysical interesting stars and mass estimates for stellar companions.
 We review  parallax results, and binary star and exoplanet mass determinations, and compare a subset of these parallaxes with preliminary \G~results. The approach to single-field relative astrometry described herein may continue to have value for targets fainter than the \G~limit in the coming era of 20-30m telescopes.
\end{abstract}




%

\section{Introduction}

With \G~poised to expand the astrometric reach of our species by many orders of magnitude \citep{Cac15}, 
this article serves to highlight the far more modest, but nonetheless  useful contributions made by
{\it Hubble Space Telescope} (\HSTs) over the past 25 years. Several Fine Guidance Sensors (FGS) aboard \HSTs have consistently delivered scientific
results with sub-millisecond of arc precision in nearly every astronomical arena to which astrometry could contribute.
These  include parallaxes for astrophysically interesting stars, both individually interesting and interesting to those in need of standard candles;
the astrometry of binary stars difficult or impossible to resolve from the ground, in aid of mass determinations; and astrometry of 
exoplanet host stars to determine inclinations with which to resolve \msini degeneracies, thus yielding true masses.

 \cite{Nel16} contains details of usage modes and figures helpful in visualizing FGS observation sequences.\footnote{http://www.stsci.edu/hst/fgs/documents/instrumenthandbook/cover.html}  An  \HST/FGS provides two types of astrometric data. Like all interferometers, an FGS produces a fringe  interrogated either by tracking the fringe central zero-crossing (POS mode) or by scanning  to build up an image of the fringe structure (TRANS mode). Parallax work rarely requires TRANS, unless the star of interest has a companion whose orbit must be established and removed. For example the binary star parallax results (e.g., Section~\ref{Bins}, below) required both POS and TRANS (see Benedict et al. 2016 and Franz et al. 1998 for details)\nocite{Ben16, Fra98} to derive a relative orbit and to measure the parallax, proper motion, and component orbits relative to reference stars. 
 
 The FGS has proven to be a versatile science instrument. For example, it has been used to observe the occultation of a star by Triton to probe that moonÕs atmosphere (Elliot et al., 1998), to measure the rotation periods of Proxima Cen and Barnard's Star (Benedict et al. 1998a), to discover sub-kilometer objects in the outer solar system (Schlichting et al, 2009, 2012), to observe light curves for transiting planets (Nutzman et al., 2011), to measure a starÕs mass using asteroseismology (Gilliland et al 2011), to resolve massive binary stars (Nelan et al., 2004, Caballero-Nieves et al., 2014, Aldoretta et al., 2015), to measure the masses of white dwarf stars (Nelan et al., 2015), and for absolute astrometry to register optical sources with radio emitters (Stockton et al., 2004, Benaglia et al., 2015).\nocite{Ell98,Sch12,Sch09,Nut11,Gil11,Nel04,Ald15,Cab14,Nel15,Sto04,Bena15, Ben98a}

This paper is  a focused review of \HST /FGS astrometry carried out to produce stellar parallaxes and exoplanet masses with some discussion of the techniques devised to produce accurate results. In Section~\ref{SGA} we compare \HST~astrometry with the global techniques of \HIP~and \G. Precision \HST~astrometry without calibration is impossible. We describe that essential calibration in Section~\ref{ROCK}. Section~\ref{WWD} summarizes  parallax (Sections~\ref{Par}, \ref{MV}, and~\ref{PR}) and exoplanet (Section~\ref{EXO}) results, the latter including both searches and companion mass determinations. In Section~\ref{GvsH} we provide a comparison of 24 of these parallax targets in common with the first \G~data release.  We briefly discuss a limited future for \HSTs astrometry in Section~\ref{Fut}. If a measure has sufficient importance,  it will eventually be improved. All of astrometry is important, as evidenced by the existence of  \G. 


\section{Single-field and Global Astrometry: \HST/FGS and \HIP} \label{SGA}
For parallax measurement the difference between single-field and global astrometry can be described by two highly simplified expressions. The \HST/FGS approach  measures only relative parallax, because both the target star of interest and the nearby (in an angular sense) reference stars have very nearly the same parallax factors. The reflex motions in the sky due to the Earth's orbital motion for both the parallax target and the surrounding reference stars are nearly identical functions of time, $f(t)$, described
\beq
f(t)*(\pi_2-\pi_1)  \rightarrow \pi_2-\pi_1
\eeq
\noindent Such measures in the absence of prior information about the reference stars produce only relative parallax. Fortunately,  \HST/FGS parallax studies have access to prior information (Section~\ref{Pr}) sufficient to yield absolute parallax.

The global astrometry produced by \HIP~(and eventually by $Gaia$) yields absolute parallax without recourse to priors,
\beq
f_2(t)*\pi_2 - f_1(t)*\pi_1 \rightarrow \pi_2 ~\textrm{and}~ \pi_1
\eeq
by simultaneously solving for the positions in two fields separated on the sky by $\sim90\arcdeg$. See \cite{Lee07a} for details of this approach, and, for example, \cite{Cac15,Soz16,Lin16} for the promise of $Gaia$.

\section{The Optical Field Angle Distortion Correction, the Bedrock of FGS Astrometry}\label{ROCK}
All \HST/FGS single-field astrometry critically depends on an Optical Field Angle Distortion (OFAD) calibration \citep{Jef94,McA02, McA06}. Precision astrometry requires corrections for optical distortions in the \HSTs Ritchey-Chretien telescope - FGS combination. The obvious solution is to map the distortions using a star field with relative positions known to better than the calibration precision. However, in the early 1990's there existed no star field with cataloged 1 millisecond of arc (hereafter, mas) precision astrometry, the desired performance goal. 

The STAT\footnote{Original STAT members included William Jefferys, P.I. and co-Investigators Fritz Benedict, Raynor Duncombe,  Paul Hemenway, Peter Shelus, and Bill van Altena, Otto Franz, and Larry Fredrick} solution was to use an FGS to calibrate itself with multiple observations at multiple offset positions and multiple roll angles of a distant star field (the open Galactic cluster, M35, d$\sim$ 830pc). A distant field insured that,
during the 2-day duration of data acquisition, relative star positions would not change. Figure~\ref{Fig-OFAD} shows the placement of the FGS pickle\footnote{We refer to the FGS field of regard as the `pickle', due to its suggestive shape.}  on the sky containing M35 cluster members. Mapping with a two dimensional fifth order polynomial and additional instrumental parameters, this self-calibration procedure reduced as-built {\it HST} telescope and FGS distortions with magnitude $\sim1\arcsec$ to below 2 mas  over much of an FGS pickle \citep{McA06}. 

The OFAD is not a static calibration. It involves changes over months and years. 
An FGS graphite-epoxy optical bench outgases for a period of time after installation in \HST. The outgassing changes the relative positions of optical components on the FGS optical bench. This results in a change in scale. 
The STAT solution to this nuisance involves revisits to the M35 calibration field periodically to monitor these scale-like changes and other slowly varying non-linearities. This is the ongoing LTSTAB (Long-Term STABility) series. LTSTABs continue as long
as it is desirable to do 1 mas precision astrometry with an FGS. See section 3.1 of \cite{Ben98} for additional details. The result of this series is to model and remove the slowly varying component of the OFAD (including, now,  proper motions to greatly improve the catalog of calibration target stars in M35), so that uncorrected distortions remain below 1 mas for center of an FGS, and below 2 mas over the entire pickle. The LTSTAB series also provides a diagnostic for deciding whether or not a new OFAD is required (for example after a significant change in \HST~due to a servicing mission).
Every astrometric result discussed in this review relies on \HSTs POS mode measures calibrated with this dynamic OFAD.

\section{A Summary of \HST /FGS Astrometric Results} \label{WWD}
The value of astrometric results can be gauged by their usage; the subsequent projects in which they have played a role. Below we summarize  initial motivations,  \HST/FGS results, and in each case, cite a few representative investigations to which they have contributed. In this review we refer to individual stars by \textbf{boldfaced number} as listed in Table~\ref{tbl-allP} and shown in the  $M_K - (V-K)_0$ HR diagram in Figure~\ref{Fig-HR}. Several of the earlier studies used STAT Guaranteed Time Observations (GTO). All others resulted from Guest Observer (GO) programs competed for on a yearly basis.

This review also serves as a roadmap of the impact that $Gaia$ will have on stellar astrophysics, the cosmic distance scale, and exoplanets with end of mission results for  thousands of times more stars of each type of interest and 10-50 times better precision than reported here.

\subsection{\HST /FGS Parallaxes}\label{Par}
 \HSTs orbits the Earth once every 90 minutes. Hence, occultations by the Earth can limit observation sequences. Depending on the ecliptic latitude of a science target and the time of year, observations can last from 40 minutes up to 90 minutes (the continuous viewing zone). During that visibility period the FGS optics bring light from the science target and (ideally) surrounding astrometric reference stars to the interferometric optics one by one in a serial fashion, cycling through the list in most cases at least twice.

In POS mode observations of each star 
usually have a duration of 60 seconds, acquiring over two thousand individual position measures. The centroid is estimated by choosing the median measure, after filtering large outliers (caused by cosmic ray hits and particles trapped by the Eath's magnetic field). The standard deviation of the measures provides a measurement error. We refer to the aggregate of astrometric centroids of each star secured during one visibility period as an ``orbit".   \HSTs astrometry is metered out in ``orbits". Multiple observations of the same science target and reference stars during an orbit allow for intra-orbit drift corrections. While these can be  corrections linear (first-order) with time, most orbits are sufficiently observation-rich to permit parabolic (second-order) corrections.

Investigators typically obtained parallax observations in pairs (typically separated by a week, a strategy originally designed to protect against unanticipated \HSTs equipment problems) at maximum parallax factors.   For most projects a few single data sets were acquired at various intervening minimum parallax factors to aid in separating parallax and proper motion. For each science target the complete data aggregate includes typically 9-11 observation sets, spanning from 1.5 to 2 yr\footnote{\HSTs astrometry only began acquiring parallax data after the first servicing mission in late 1993. This delay did not involve the mis-figured primary mirror, but rather the original solar panels. These incorporated a flawed design that caused excessive shaking of the entire telescope, often with amplitudes sufficient to knock the FGS out of fine lock on the fringe zero-crossing. The replacement panels solved this problem \citep{Ben98}.}. Investigators prefer to place the science target in the center of the FGS pickle, where the OFAD applies the best correction. Central placement is not always possible, because of guide star availability in the other two FGS, and, due to Sun constraints, \HSTs will have rolled 180$\arcdeg$ from one maximum parallax factor to the next, six months later.

\subsubsection{Single-Field Priors}\label{Pr}
The success of single-field parallax astrometry depends on prior knowledge of the reference stars, and sometimes, of the science target. Catalog proper motions with associated errors, lateral color corrections, and estimates for reference star parallax are entered into the modeling as quasi-Bayesian priors, data with which to inform the final solved-for parameters. These values are not entered as hardwired quantities known to infinite precision. They are observations with associated errors. The model adjusts the corresponding parameter values within limits defined by the data input errors to minimize $\chi^2$.

\textbf{Reference Star Absolute Parallaxes-} Because the parallaxes determined for the targets listed in Table~\ref{tbl-allP} are measured with respect to 
reference frame stars which have their own parallaxes, investigators must either apply a statistically-derived correction from relative to absolute parallax \citep[Yale Parallax Catalog, YPC95]{WvA95}, or 
estimate the absolute parallaxes of the reference frame stars. The latter is the approach most investigators have adopted since  first used  in \cite{Har99}. 
The colors, spectral type, and luminosity class of a star can be used to estimate the 
absolute magnitude, $M_V$, and $V$-band absorption, $A_V$. The absolute parallax for each reference star is then,
\begin{equation}
\pi_{\rm abs} = 10^{-(V-M_V+5-A_V)/5}
\end{equation}

Band passes for reference star photometry can include: $BVRI$ and $JHK$ (from 2MASS\footnote{The Two Micron All Sky Survey
is a joint project of the University of Massachusetts and the Infrared Processing
and Analysis Center/California Institute of Technology }). Parallax work often has access to spectroscopy with which to estimate stellar spectral type and luminosity class. Classifications used a combination of template matching and line ratios.  
Absorptions are estimated from a combination of $A_V$ maxima \citep{Schl98} and  comparing colors with spectroscopic analysis.
The reference star derived absolute magnitudes are critically dependent on the assumed stellar 
luminosity, a parameter impossible to obtain for all but the latest type stars 
using only  color-color diagrams. To confirm the luminosity classes obtained from classification spectra (or for those targets with no available reference star spectra) users often gather  previously measured proper 
motions for a  field centered on the parallax target, and then 
iteratively employ the technique of reduced proper motion \citep{Yon03,Gou03}  to discriminate 
between giants and dwarfs.  Reference star absolute parallaxes are derived by comparing  estimated spectral types and luminosity 
class to  $M_V$ values from \nocite{Cox00} Cox (2000). 

Adopted input errors for reference star
distance moduli, $(m-M)_0$, are typically 1.0 mag.  Contributions to the error are uncertainties in $A_V$ and errors 
in $M_V$ due to uncertainties in color to spectral type mapping.   Typically, no reference star absolute parallax is better determined than $\frac{\sigma_{\pi}} {\pi}$ = 
23\%. For example, an average input absolute parallax for the reference frame might be 
$\langle\pi_{abs}\rangle = 1.0$ mas. Comparing this to the correction to absolute parallax discussed and presented in 
YPC95 (section 3.2, figure 2), entering YPC95, figure 2, with a Galactic latitude, $b = 
-40\arcdeg$, and average magnitude for the reference frame, $\langle V_{\rm ref} \rangle 
= 14.85$, one obtains a correction to absolute of 1.2 mas, consistent with the  derived correction.

\textbf{Proper Motions-} For example \cite{Ben07, McA14} use proper motion priors from the  PPMXL \citep{Roe10}, UCAC4 \citep{Zac13}, or URAT \citep{Zac15} catalogs. These quantities typically have errors on order 
4 mas yr$^{-1}$.

\textbf{Lateral Color Corrections-}  To effectively periscope the entire pickle, the FGS design included refractive optics. Hence, a blue star and a red star at exactly the same position on the sky would be measured to have differing positions within the pickle, a correctable offense. The discussion in section 3.4 of \cite{Ben99} describes how to derive this correction (lc$_x = -0.9\pm0.2$ mas, lc$_y = -0.2\pm0.3$ mas) for FGS\,3. A similar 
process resulted in FGS\,1r lateral color corrections (lc$_x = -1.1\pm0.2$ mas, lc$_y = -0.7\pm0.1$ mas), all quantities introduced as observations with error in the models.

\textbf{Crossfilter Corrections-} One of the significant strengths of the FGS for astrometry is the dynamic range of the device, permitting us to relate the positions of bright targets to a far fainter set of astrometric reference stars. For example \citep{Ben07} the Cepheid $\ell$ Car (\textbf{1}) has  $\langle V \rangle = 3.73$ and a reference star field characterized by  $\langle V_{\rm ref} \rangle =13.00$. FGS\,3 and FGS\,1r are equipped with filter wheels containing neutral density filters providing 5 magnitudes of attenuation. No filter has perfectly plane-parallel faces. This filter wedge introduces a position shift when comparing filter-in relative to filter-out. Section 2 of \cite{Ben02a} details this calibration process for FGS\,3, a process subsequently carried out for FGS\,1r.

\subsection{The Astrometric Model}
All prior information now enters the modeling as observations with error. From  
positional measurements of the reference stars one determines the scale, rotation, and offset ``plate
constants" relative to an arbitrarily adopted constraint epoch for
each observation set. Most \HST /FGS astrometry employs GaussFit (Jefferys \etal 
1988) 
\nocite{Jef88} to minimize $\chi^2$. The solved equations of condition for  
a typical
field are
\begin{equation}
        x^\prime = x + lc_x(\it B-V) ~[- \Delta XF_x] 
\end{equation}
\begin{equation}
        y^\prime = y + lc_y(\it B-V)  ~[- \Delta XF_y] 
\end{equation}
\begin{equation}
\xi = Ax^\prime + By' + C  - \mu_\alpha \Delta t  - P_\alpha\pi ~[- (ORBIT_{b,x} + ORBIT_{c,x})]
\end{equation}
\begin{equation}
\eta = Dx^\prime + Ey^\prime + F  - \mu_\delta \Delta t  - P_\delta\pi ~[- (ORBIT_{b,y} + ORBIT_{c,y})]
\end{equation}

\noindent 
where square brackets, [ ... ] indicate optional terms included only for some models. Identifying terms, $\it x$ and $\it y$ are the measured FGS coordinates from {\it HST};   $(B-V)$ is the Johnson $(B-V)$ color of each star; $\it lc_x$ and $\it lc_y$ are the lateral color corrections; $\Delta XF_x$ and $\Delta XF_y$ are the cross-filter corrections.   $A$, $B$, $D$ and $E$ are scale and rotation plate constants, $C$ and $F$ are offsets; $\mu_\alpha$ and $\mu_\delta$ are proper motions; $\Delta t$ is the epoch difference from the constraint epoch; $P_\alpha$ and $P_\delta$ are parallax factors;  and $\it \pi$ is  the parallax.   It is necessary to obtain the parallax factors from, for example,  the JPL Earth orbit predictor 
(Standish 1990)\nocite{Sta90}, version DE405. Orientation to the sky is obtained from ground-based reference star astrometry (UCAC4 or PPMXL Catalogs) with uncertainties 
of $0\fdg05$. Note that for most target field investigations $D=-B$ and $E=A$, constraining equality for the scales along each axis.

The optional terms $ORBIT_x$ and $ORBIT_y$ are functions of the classic binary star parameters; semi major axis, inclination, eccentricity, argument of periastron, longitude of ascending node, orbital period, and time of periastron passage \citep{Hei78,Mar10}. These terms are included for binary star astrometry (e.g. Benedict et al. 2016)  and exoplanet perturbation studies (e.g. McArthur et al. 2014).
An FGS provides positions of the host star (for exoplanet host measures) or stars (for binary star studies) on the plane of the sky at different epochs. 
Equations 6 and 7 describe a model with an option to characterize a perturbation caused by a two planet system with components b and c.

\subsection{Absolute Magnitudes} \label{MV}
\HST /FGS parallax efforts have impacted two major arenas: distance scale and stellar astrophysics. Absolute magnitudes become the primary goal for every campaign. Absolute magnitudes, M, are related to parallax through
\beq
(m-M) = -5 \rm log(\pi_{\rm abs}) -5 +A_V
\eeq
\noindent where m is apparent magnitude, (m-M) the distance modulus, $\pi_{\rm abs}$ is at the measured absolute parallax in arc seconds, and A$_V$ the absorption, in this case in the $V$ band. Over the years investigators have produced over 100  parallaxes and  absolute magnitudes (Table~\ref{tbl-allP}). 

Figure~\ref{Fig-HR} presents a Hertzsprung-Russell (HR) diagram using only absolute magnitudes, $M_K$, generated from the parallax results presented in Table~\ref{tbl-allP}. This diagram  displays over 18 magnitudes of intrinsic luminosity range, from  GJ 1245C (\textbf{48} in Table~\ref{tbl-allP} and Figure~\ref{Fig-HR}), a very late M dwarf just at the brown dwarf-star mass separation, to  $\ell$ Car (\textbf{1} in Table~\ref{tbl-allP} and Figure~\ref{Fig-HR}), the longest period Cepheid in that study. The choice of $K$-band reduces the impact of errors in establishing the absorption, $A$, $A_K$ having nine times less effect than $A_V$ \citep{Schl98}. The absorption can often be a significant source of uncertainty. Note that the extremely blue and intrinsically faint white dwarf, RE\,J\,0317-853 (\textbf{104}), was not plotted ($M_K=13.12, (V-K)(0)=-0.79$) to preserve
a readable separation of ID numbers along the y-axis. Thus, Table~\ref{tbl-allP} encompasses over 20 magnitudes in absolute $K$ magnitude. 

Bias questions arise when considering measured quantities involving distance. Lutz \& Kelker (1973) \nocite{Lut73} used a frequentist argument to show that, if stars of the
same measured parallax are grouped together, the derived mean
parallax will be overestimated. This is because for most Galactic
stellar distributions, the stars with overestimated parallaxes will
outnumber those with underestimated parallaxes. \cite{Ben07,Ben09,Ben11,Roe07} use the subsequent 
modification by Hanson (1979), thus an LKH correction\nocite{Han79}. They applied this bias correction to single stars by putting the
argument in a Bayesian form.

Invoking Bayes Theorem to assist with generating absolute magnitudes from  parallaxes, one would say, ``what is the probability that a star from this population with this position would have parallax, $\pi$, given that we have not yet measured $\pi$?" One must use the space distribution of the population to which the star presumably belongs. This space distribution is built into the prior p($\pi$) for $\pi$, and is used to determine
\beq
 p(\pi | \pi_{observed} \& K) \sim p(\pi_{observed} | \pi \& K) p(\pi | K)
\eeq
\noindent where K represents prior knowledge about the space distribution of the class of stars in question. The ``\&" is an ``and"  operator. The function p($\pi_{observed} | \pi$ \& K) is the
standard likelihood function, usually a gaussian normal with variance $\sigma_{\pi}$. The
"standard" L-K correction has p($\pi |$ K) $\sim\pi^{-4}$. Looking at a
star in a disk population close to the galactic plane requires $\pi^{-3}$
(ignoring spiral structure), which is the prior  typically used.
The LKH bias is proportional to $(\sigma_{\pi}/\pi)^2$. For example the average LKH bias correction for the Cepheids (\textbf{1-10})  and the RR Lyrae stars (RRL, \textbf{13 -- 17}) in Table~\ref{tbl-allP}  was -0.06 magnitude. No LKH bias corrections were needed for the nearby M dwarfs (\textbf{31-54}). Lastly, note that in \cite{Ben11} as a check of the LKH correction the final RRL zero-point was also obtained through an entirely different  bias correction approach, that of reduced parallaxes \citep{Fea02}, resulting in an RRL $M_V$ at [Fe/H]=-1.5 that differed by 0.01 magnitude from that derived with LKH bias corrections.

The scatter along the Main Sequence in Figure~\ref{Fig-HR} cannot be due to either parallax error (typically 1\% or 0.01 magnitude) or reddening estimation uncertainties. Most of those stars have an absorption $A_K < 0.05$. Therefore that scatter is intrinsic and offers multiple opportunities to test the ability of stellar models to account for age, metallicity, convection, turbulent mixing, rotation, mass loss, and magnetic fields \citep{And91}. 

\subsection{Parallax Results} \label{PR}
\subsubsection{Standard Candles}
Cepheids and RRL play a central role in the cosmic distance scale. Both stellar types exhibit a behavior consistent with a Period-Luminosity relationship (PLR), the Leavitt Law. Once located, and once periods are in hand, comparison of aggregates of these in some distant galaxy or globular cluster will yield a PLR that can be compared to a calibrated PLR to provide magnitude offset, hence, distance. The proposal to measure the parallaxes of ten Galactic Cepheids was the top rated proposal in the  2003 competition for \HSTs time\citep{Ben04}. Subsequently, \cite{Ben07} provided parallaxes with an average error of 8\% for ten Galactic Cepheids (\textbf{1 - 10}). Uses have ranged from distance scale refinements \citep{Bha16,Mus16} to detailed investigations of Cepheid astrophysics \citep{Bre16,And16}.

\cite{Ben11} measured the parallaxes of five RRL (\textbf{13 -- 17}) and two dwarf Cepheids (\textbf{11 - 12}). These few objects (with a 6.7\% average parallax precision) could not yield  independent slopes for either PL or luminosity-metallicity relationships, but did provide zero-points, applicable to Galactic globular cluster distance determinations \citep{Nee15} and RRL astrophysics \citep{Bre15}. The \HSTs /FGS parallax for VY Pyx placed this star over a magnitude below a $K-$band Period-Luminosity relation comprised of the five RRL and another Type II Cepheid, $\kappa$ Pav. To date, no subsequent investigations have explained this astrophysically interesting aberration (although see Section~\ref{GvsH}, below).

Figure~\ref{Fig-PLR} contains the \cite{Ben07} Cepheid $K$-band Leavitt Law and the PLR constructed from the \cite{Ben11} RRL and dwarf Cepheid parallaxes. 

\subsubsection{Galactic Clusters}
Since the publication of the original \citep{Per97} and revised \citep{Lee07a} \HIP~catalogs, tension existed between the \HIP~significantly smaller distance ($\sim$125pc) to the Pleiades and literally every other determination by multiple methods, including the \HST /FGS result (\textbf{28 - 30}, Soderblom et al. 2005, 135.6 $\pm$ 3.1pc)\nocite{Sod05}. Recently very long baseline interferometry has provided a parallax (Melis et al. 2014, 136.2$\pm$1.2 pc)\nocite{Mel14} agreeing with \cite{Sod05} within the error bars. Even more recent but preliminary results from \G~ are completely consistent with both those determinations (Section~\ref{GvsH}, below).

The Hyades is another cluster useful for both distance scale and stellar astrophysics. \HST/FGS has worked on this cluster twice, the first time with a preliminary OFAD and a more primitive analysis approach \citep{WvA97b}. The refined distance modulus derived from parallax measurements of seven Hyads (\textbf{19 - 25}, McArthur et al. 2011) \nocite{McA11}, m-M= 3.38$\pm$0.01, compares with \HIP, m-M=3.33$\pm$0.02 \citep{Lee07a}. The disagreement could involve  sparse (compared to \HIP) sampling of cluster membership.

\subsubsection{Planetary Nebulae Central Stars}

\cite{Ben09} obtained parallaxes of four planetary nebulae central stars (CSPN, \textbf{56 - 59}), motivated by a lack of agreement among more indirect methods of distance determination, and a desire to assist in the calibration of those indirect methods. For this investigation the final parallaxes were a weighted average with previous determinations \citep{Har07}, yielding an average precision of 7.6\%. \cite{Smi15, Ali15} have used these results to further explore inconsistencies among the indirect distance determination techniques, including expansion and spectroscopic parallax methods.
Other studies of individual objects have benefitted from direct distance measures \citep{Lag15, Ste15}. 

\subsubsection{The AM CVn Stars}
AM CVn stars are white dwarfs (WDs) accreting matter from a degenerate or semidegenerate companion. Overviews of this class of ultracompact binary stars are given by Warner (1995)\nocite{War95}, Nelemans (2005)\nocite{Nel05}, and \cite{Sol10}. The \HST/FGS combination provided five parallaxes (\textbf{64 - 68}) with an average 9.4\% precision \citep{Roe07}. These parallaxes, allowing the determination of absolute energetics, have anchored theoretical explorations of outbursts \citep{Kot12} and predictions of gravitational wave amplitudes \citep{Lop14}.

\subsubsection{Cataclysmic Variables and Novae}

\HST /FGS has measured parallaxes for nine cataclysmic variables (CVs, \textbf{69 - 77}, McArthur et al. 1999, 2001; Beuermann et al. 2003, 2004; Harrison et al. 2004; Nelan et al 2013, Harrison and McArthur 2016)\nocite{McA99,McA01,Nel13} \nocite{Har99} \nocite{Beu03, Beu04} \nocite{Har16} and of four classical novae (\textbf{60 - 63}, Harrison et al. 2013)\nocite{Har13}. Precise distances aid the  exploration of the reasons for intrinsic variations among the various subclasses of cataclysmic variables. See \cite{Har14} for a recent overview. Related in a sense to this group is Feige 24 (\textbf{55}, Benedict et al. 2000)\nocite{Ben00a}, a WD - M dwarf binary described as the prototypical post common-envelope detached system with a low probability of becoming a cataclysmic variable (CV) within a Hubble time. This object was selected for the STAT original Guaranteed Time Observing  parallax program because a directly measured distance could reduce the uncertainty of the radius of one of the hottest white dwarfs. Thus far this parallax aided in better determinations of the Feige 24 WD and M dwarf masses and system inclination \citep{Kaw08}.

Tables~\ref{tbl-allP} and \ref{tbl-HG} list two entries for SS Cyg (\textbf{70, 70A}). The difference in parallax value results from difference in the treatment of the astrometric reference frame. For example \cite{Har16} used all available reference stars, \cite{Nel13} removed one. There were also differing choices made regarding reference star priors. This highlights one of the major potential pitfalls of single-field relative astrometry, a limited number of nearby reference stars and sensitivity to problems with any. It should be pointed out that this issue seldom arises. Most investigators run multiple models, removing single reference stars in turn to assess
any problems, and rarely find any.

\cite{Kub10} presented FGS parallaxes (\textbf{104 -105}) for the massive, hot, and rapidly rotating WD, RE\,J 0317-853 and its companion LB9802.

\subsubsection{M Dwarf Binary Stars} \label{Bins}
With the exception of the Hertzsprung-Russell diagram, the Mass-Luminosity Relation (MLR) is  the single most important ``map" in stellar astronomy (e.g., Henry et al. 1999). \nocite{Hen99}  The entire evolution of a star depends on its mass. 
The MLR allows astronomers to convert a $relatively$ easily observed quantity, luminosity, to a more revealing characteristic, mass. This provides a better understanding of the star's nature. M dwarfs make up at least 75\% of all nearby stars \nocite{Hen06}(Henry et al. 2006). An accurate MLR permits a luminosity function to be converted to a mass function, and drives estimates of the stellar contribution to the mass of the Galaxy. Additionally, precise masses challenge models of low-mass stellar evolution.

For binary stars TRANS mode yields the structure of the fringe from which it is possible to derive the separation and position angle of the primary component  relative to the secondary component. This establishes the relative orbit. POS mode permits the measurement of the position of the primary star (brighter component) relative to a local frame of reference. This   determines proper motion, \textbf{parallax}, and mass fraction. Binary component mass precision depends critically on parallax precision.

Parallaxes are essential to the successful completion of this, one of the longest-running \HSTs projects \citep{Hen95}, involving data collection from 1994 to 2009. The need to monitor the orbital motion of long-period binaries drove the study duration. A final paper (Benedict et al. 2016) presents $V$ and $K$-band MLR\nocite{Ben16}. They list new and improved parallaxes for the components of twelve systems (\textbf{31 - 54}) with an average precision of 0.4\%. These are used to produce the $K$-band MLR shown in Figure~\ref{Fig-MLR}.

\subsubsection{Metal-poor Main Sequence Stars}
\cite{McC97} reported parallaxes for two high-velocity stars (\textbf{78 - 79}). While this early effort yielded lower precision than typically achieved with the FGS,
the parallaxes nonetheless  identify them as sub-dwarfs (Figure~\ref{Fig-HR}). \cite{Bon13} and \cite{VdB14} added parallaxes for three more metal-poor objects off the main sequence (\textbf{100 - 103}). Investigators have more recently obtained parallaxes of eight metal-poor ([Fe/H] $< -1.5$) main sequence stars (\textbf{80 - 87}, Chaboyer et al. 2016, in prep). These  have an average precision of 1.3\%, providing  absolute magnitude uncertainties of $\le0.05$ mag for a given star. These stars will be used to test metal-poor stellar evolution models and to determine more precise main sequence fitting distances to a large number of low metallicity globular clusters.

\subsection{A First Look at \G~Results and a Comparison with \HST /FGS} \label{GvsH}
During the final preparation of this review, the \G~team released a preliminary catalog (Data Release 1, hereafter, \G~DR1) containing the five classical astrometric parameters, position in RA and DEC, proper motion in RA and DEC, and parallax for a subset of the total number of stars ultimately to be measured \citep{deB16,Bro16,Fab16,Lin16,Cro16}.  In contrast to the final \G~catalog, \G~DR1 does involve prior information from the Tycho catalog. Table~\ref{tbl-HG} lists 26 objects found in \G~DR1 with \HSTs parallaxes. Values from the \cite{Lee07a} \HIP~ reanalysis are also listed, along with the mean and median parallax error for each mission. Figure~\ref{Fig-GH} shows an impartial regression line and residuals derived from a GaussFit model (Jefferys et al. 1988) that fairly assesses errors in both \HSTs and \G~parallaxes. See \cite{McA11} for a similar regression between \HSTs and \HIP~parallaxes. For \G~and \HSTs parallaxes we find no significant scale difference over a parallax range $2 <\pi_{\rm abs} <  40$ mas. The median error for the \HIP~parallax measures is 1.31 mas, for \G~DR1 0.33 mas\footnote{not including an estimated systematic error of 0.3 mas, as stated in http://www.cosmos.esa.int/web/gaia/dr1}, and for the \HSTs measures, 0.17 mas. A linear fit ({Figure~\ref{fig-GHV}) of $\Delta \pi_{abs} (\G-\HST)$ against V magnitude has a Pearson's r value of -0.34, indicating a weak correlation. 
We see a trend of larger differences for brighter stars. Future \G~releases should yield parallaxes far more precise than \HST. 

\G~DR1 contains only one Cepheid (FF Aql, \textbf{5}), no M dwarf binaries, and no CSPN in common with  Table~\ref{tbl-allP} targets. \G~DR1 does include VY Pyx (\textbf{12}), with a parallax that brings it much closer to the RRL and dwarf Cepheid PLR in Figure~\ref{Fig-PLR}. Interestingly, the one Pleiades star in common (P3179, \textbf{28}) has a parallax closer to the \HIP~ Pleiades value than to the \HSTs value \citep{Sod05}. However, the distribution of all Pleiades star parallaxes in \G~DR1 has a mean \citep{Bro16}  agreeing with \HSTs \citep{Sod05}.

\subsection{Exoplanetary Searches and Masses}\label{EXO}
The past 20 years have seen an explosion in the study of extrasolar planets. The discovery of thousands of planets orbiting other stars has captured the imagination of the astronomical community and the public at large. Many surprises have popped up along the way, such as the discovery of Jupiter-mass planets in orbits with periods less than 10 days and a terrestrial-sized planet in the temperate zone of the M dwarf, Proxima Cen \citep{Ang16}. These discoveries have challenged and modified our understanding of planet formation and the importance of planetary migration. For reviews of the properties of known extrasolar planets and implications for planet formation see for example Udry et al. (2007), Lissauer \& Stevenson (2007), Durisen et al. (2007), Papaloizou et al. (2007),  \cite{Per11}, \cite{Win15AR}, and \cite{Bor16}. \nocite{Udr07} \nocite{Lis07} \nocite{Dur07} \nocite{Pap07}

Before $Kepler$ \citep{Bor10,Bor16} flooded us with exoplanets, and before precision radial velocities started that flood, e.g.,  \cite{But06}, the question of their existence was sufficiently compelling that beginning in 1992  \HSTs astrometrically monitored two stars, first Proxima Cen and later Barnard's Star (\textbf{26 - 27}), in a detection effort utilizing fifty-nine and thirty-four orbits respectively. This large number of orbits for a search campaign included those from  early STAT GTO Science Verification studies using the Proxima Cen field.  \cite{Ben99} reported  null results (detection lower limits only). The only differences between a detection and a parallax campaign are number of observations, their cadence, the duration, and the forbearance of the Time Allocation Committee.

Other systems became targets for \HSTs astrometry, because of prior knowledge of (possibly) substellar companions, usually from precision radial velocity (RV) investigations. Given that time on \HSTs is extremely difficult to obtain, RV data proved invaluable in every case, phasing the \HSTs astrometry measures (often sampling less than a full companion period), permiting detection and characterization of reflex astrometric motion. Table~\ref{tbl-allM} summarizes the results of  \HST/FGS studies of exoplanet host stars to determine companion masses. Every target on that list had a known companion, with a known (usually large) \msini, and a period and estimated parallax that predicted a detectable perturbation of the star position due to the companion. 

These \HSTs data might ultimately be combined with \G~ measures, to extend significantly  the time baseline of astrometry, thereby improving proper motion and exoplanet perturbation characterization.

We now discuss each of the past exoplanet mass results in turn, starting with the earliest investigation. Boldface \textbf{(numbers)} refer to parallax results listed in Table~\ref{tbl-allP} and identify the host star in the Figure~\ref{Fig-HR} HR diagram.

\begin{enumerate}
\item \textbf{GJ 876 (88) - } In 1998, two groups (Delfosse et al. 1998; Marcy et al. 1998) \nocite{Del98, Mar98}  announced the discovery (with RV) of a companion to this M4 dwarf star, Ross 780 = Gl 876. The companion was characterized by \msini $\sim$ 2 \mjup and P $\sim$ 60 days. A second companion was later detected, Gl 876c \nocite{Mar01} (Marcy et al. 2001), with \msini $\sim$ 0.56 \mjup and P $\sim$ 30 days. The \HSTs TAC dedicated \HSTs time to measure the perturbation due to the longer period, more massive companion, Gl 876b. The low mass of the primary (\m$_*$ = 0.32 \msune), the period of Gl 876b, and the system proximity (d = 4.7 pc) suggested that even an edge- on orientation ($\cal{M} =$ \msini) would produce a detectable perturbation, thus a companion mass. A total of twenty seven orbits with \HSTs FGS\,3 resulted in \m$_{b}$ = 1.89$\pm$0.34\mjupe, the first astrometrically determined mass of an extrasolar planet \citep{Ben02c}.

\item \textbf{$\rho^1$Cnc (93) - } \cite{McA04} announced the discovery (using RV from multiple sources) of $\rho^1$ Cnc e, a short-period fourth addition to this already rich exoplanetary system. Archived \HSTs FGS\,3 astrometry data, phased with the RV data  yielded a measurement of the inclination of the outermost companion, component  d, thus its mass, \m$_d$ = 4.9$\pm$1.1\mjupe.  A re-analysis of existing RV data yielded a shorter period for component e, $P_e = 0.74$d \citep{Daw10}. Subsequently, component e  was shown to transit its host star \cite{Win11}.

\item \textbf{$\epsilon$ Eri (95) - } Many RV observations existed for this young, spectroscopically active star. An exoplanetary origin for the 
P = 6.85 y RV signature has not been universally agreed upon \citep{Met13}. Observational material consisted of 46 orbits of FGS\,1r data collected over three years,  130 epochs (secured over 14 years) of ground-based MAP astrometry \cite[for example]{Gat06}, and 235 RV measures spanning over 24 years. Simultaneous modeling yielded a perturbation orbit due to $\epsilon$ Eri b. Assuming a stellar mass, \m$_*$= 0.83\msune, \cite{Ben06} obtained  \m$_{b}$ = 1.55$\pm$0.24\mjupe. Multiple attempts (summarized in Bowler 2016\nocite{Bow16}) have yet to yield a direct detection of $\epsilon$ Eri b.

\item \textbf{HD 33636 (97) - } This was considered  a good candidate for \HSTs astrometry due to its proximity \citep{Per97}, confirmed by \HST /FGS (Table~\ref{tbl-allP}), and large \msini \citep{But06}. Analysis (identical to that for $\epsilon$ Eri) combined 18 orbits of \HSTs FGS\,1r astrometry and over 140 RV measures from four sources to yield a surprise; the companion has a mass  \m$_{B}$ = 142$\pm$11\mjupe = 0.14$\pm$0.01\msune. It is an M dwarf star \citep{Bea07}. Note that there exists a fairly large parallax disagreement, comparing to \G~ (Table~\ref{tbl-HG}), possibly because neither \HSTs nor \G~ fully sampled the 5.8 year period of this binary system.

\item \textbf{$\upsilon$ And (89) - } Over 4.5 years \cite{McA10} obtained fifty-four orbits of \HSTs FGS\,1r data for this dynamically interesting, nearby,  multi-planet system \citep{But99} with a goal of determining the masses of two of the companions, components c and d. Simultaneous modeling of \HSTs astrometry and 974 RV measures spanning 13 years yielded \m$_{c}$ = 13.98$\pm$3.8\mjup and \m$_{d}$ = 10.25$\pm$2\mjupe, and another surprise. The system is not coplanar, having a mutual inclination of $\Phi =29\fdg9\pm1\fdg0$ \citep{McA10}. These results continue to engage those interested in planetary system formation, architectures, and lifetimes (c.f. Deitrick et al. 2015).\nocite{Dei15}

\item \textbf{HD 136118 (96) - } Found (by RV) to host an exoplanet (\msini$_b$ = 11.9\mjupe) by Fischer et al. (2006)\nocite{Fis02}, and relatively nearby \citep{Per97},  18 orbits of \HSTs FGS\,1r astrometry were secured, which, when combined with RV data resulted in \m$_b=42^{+11}_{-18}$\mjupe ~\citep{Mar10}. HD 136118 b remains one of the few brown dwarfs with an astrometrically measured mass \citep{Wil16}.

\item \textbf{HD 38529 (90) - } This star hosts two known companions discovered by high-precision RV monitoring (Fischer et al. 2001, 2003; Wright et al. 2009) \nocite{Fis01,Fis03,Wri09} with minimum masses \msini$_b$ = 0.85 \mjupe and \msini$_c$ = 13.1 \mjupe, the latter right above the currently accepted brown dwarf mass limit. A predicted minimum perturbation for the outermost companion, HD 38529c, $\alpha_c = 0.8$ mas, motivated us to obtain  \HSTs FGS\,1r astrometry over 3.25 yr with which to determine the true mass. Analysis of RV from four telescopes with over ten years coverage and 23 orbits of \HSTs FGS\,1r astrometry resulted in $\alpha_c = 1.05 \pm 0.06$ mas and \m$_c = 17.6^{+1.5}_{-1.2}$\mjup \citep{Ben10}, 3$\sigma$ above a 13\mjup deuterium burning, brown dwarf lower limit. That modeling included a ``d" component to lower RV residuals,  very tentatively interpreted as  a potential companion. \cite{Ben10} suggested caution in accepting that interpretation, caution later vindicated by the erasure of component d with additional high-precision RV \citep{Hen13}. 

\item \textbf{HD 128311 (94) - } Yet another multi-planet system discovered with RV \citep{Butl03, Vog05}, the same team involved with the  $\upsilon$ And investigation secured 29 orbits of FGS\,1r astrometry in an effort to establish system architecture. The analysis \citep{McA14} resulted in an inclination and mass only for component c, $i_c = 56 \pm 14$\arcdeg~with \m$_c = 3.8^{+0.9}_{-0.4}$\mjupe. Given  demonstrated perturbation sensitivity \citep{Ben02c}, the inferred inclination for component d with \msini=0.13\mjup is $i_d > 30$\arcdeg. An entire cottage industry has built up around the analyses of mean motion resonances (MMR) in this system. Joining the fray, \cite{McA14} argued against MMR, based on the RV data. The arguments continue \citep{Rei15}.

\item \textbf{XO-3 (98) - } 
Johns-Krull et al. (2008) \nocite{Joh08} announced the discovery of a massive ($\sim$13 \mjupe) planet (XO-3b) in an eccentric ($e \sim$ 0.22), 3.2 d transiting orbit around an F5V star. For this system a combination of transits and RV yield mass. They found evidence that the planet may be  larger (R$_P \sim 1.9 R_{Jup}$) than expected,  inconsistent with standard structure models (e.g., Fortney \& Nettlemann, 2010) \nocite{For10}which include irradiation by the central star. That radius suggests the need for additional internal heating in XO-3b, perhaps tidal, due to the eccentric orbit (e.g., Gu et al. 2003, Adams \& Laughlin et al. 2006).\nocite{Gu03} \nocite{Ada06} Determining the planetary parameters requires knowledge of the stellar parameters, and the planetary radius is particularly sensitive to these. \HST /FGS parallax precision  provides a more precise determination of host star characteristics, hence the radius of XO-3b (Johns-Krull 2016, in preparation).


\end{enumerate}

We expect these early exoplanet mass results to increase in number substantially once the final $Gaia$ data releases are made available \citep{Cas08,Per14,Soz16}.

\section{The Future}\label{Fut}
The first promise that \G~delivered on was to shut down \HST/FGS POS mode astrometry, long before \G~launch and $any$ evidence of success. Only two FGS POS mode proposals have succeeded in getting time since 2011 (a total of eight orbits).

The parallax and exoplanet mass results summarized in this review required more than 1500  \HSTs orbits over 25 years, less than 1\% of the total allotted orbits. \HSTs has a few more contributions to make to precision astrometry. \nocite{Ker14} Kervella et al. (2014) are measuring the parallax of the classical Cepheid, V1334 Cyg. Funding issues have delayed the final analysis  of FGS\,1r  measures of the exoplanetary systems $\mu$ Ara, HD 202206,  and $\gamma$ Cep, with results likely available next year. Beyond that time frame, \HSTs will continue to produce new astrometry, but probably not with the FGS. 
\cite{Ries14} and \cite{Cas15} report 20-30 microsecond of arc parallaxes, using drift scanning with Wide Field Camera 3. This  program  will extend the Galactic Cepheid PLR to longer periods and densify the existing calibration with many more Cepheids. Also, should \G~bright star astrometry fall short, there remain a few exoplanet host stars suitable for drift scan astrometry  to establish companion true mass.

Additionally, these single-field techniques might be useful to future astrometric users of, for example, the Large Synoptic Survey Telescope (Ivezic et al. 2008)\nocite{Ive08}, or the Giant Magellan Telescope \citep{McC14} when they require the highest-possible astrometric precision for targets of interest (fainter than \G~limits) contained on a single CCD in the focal plane.

\acknowledgments

The original Space Telescope Astrometry Team, formed in late 1977, consisted of  William Jefferys, P.I. and co-Investigators Fritz Benedict, Raynor Duncombe (deceased),  Paul Hemenway, Peter Shelus (all University of Texas), and Bill van Altena (Yale University), Otto Franz (Lowell Observatory), and Larry Fredrick (University of Virginia, and coiner of the 'pickle' moniker for the FGS field of regard). They immediately augmented the STAT with the essential participation of Barbara McArthur (University of Texas), Ed Nelan (STScI), Darrel Story (GSFC), Larry Wasserman (Lowell Observatory), Phil Ianna (University of Virginia) and Terry Girard (Yale University). While the STAT officially ceased operations in 1998, their continued enthusiasm, support, and scientific acumen touched every aspect of the work reviewed herein. Support for this work was provided by NASA through grants 2939, 2941, 3004, 3061, 3886, 4031, 4758, 4884, 4892, 4893, 4935, 4938, 5054, 5056, 5067, 5174, 5586, 5587, 5657, 6036, 6037, 6047, 6157, 6158, 6238, 6239, 6240, 6241, 6262, 6263, 6264, 6267, 6268, 6269, 6270, 6538, 6566, 6764, 6768, 6873, 6874, 6875, 6877, 6879, 6880, 7492, 7493, 7894, 8102, 8292, 8618, 8729, 8774, 8775, 9089, 9167, 9168, 9190, 9230, 9233, 9234, 9338, 9347, 9348, 9407, 9407, 9408, 9879, 9879, 9969, 9971, 9972, 10103, 10104, 10106, 10432, 10610, 10611, 10613, 10704, 10773, 10929, 10989, 11210, 11211, 11299, 11704, 11746, 11788, 11789, 11942, 12098, 12320, 12617, 12629, and 12629
 from the Space Telescope 
Science Institute, which is operated
by the Association of Universities for Research in Astronomy, Inc., under
NASA contract NAS5-26555. This publication makes use of data products from the 
Two Micron All Sky Survey, which is a joint project of the University of 
Massachusetts 
and the Infrared Processing and Analysis Center/California Institute of 
Technology, 
funded by NASA and the NSF.  This research has made use of the {\it SIMBAD} and {\it Vizier} databases, 
operated at Centre Donnees Stellaires, Strasbourg, France; Aladin, developed and maintained at CDS; the NASA/IPAC Extragalactic Database (NED) 
which is operated by JPL, California Institute of Technology, under contract 
with 
NASA;  and NASA's truly essential Astrophysics Data System Abstract Service. This work has made use of data from the European Space Agency (ESA)
mission {\it Gaia} (\url{http://www.cosmos.esa.int/gaia}), processed by
the {\it Gaia} Data Processing and Analysis Consortium (DPAC,
\url{http://www.cosmos.esa.int/web/gaia/dpac/consortium}). Funding
for the DPAC has been provided by national institutions, in particular
the institutions participating in the {\it Gaia} Multilateral Agreement. Many people over the years have materially improved all aspects of the work reported, particularly Linda Abramowicz-Reed, Art Bradley, Denise Taylor, and all the co-authors of our many papers. 
G.F.B. thanks Debbie Winegarten, whose able assistance with other matters freed me to devote necessary time to this Review. We thank an anonymous referee for their careful review that resulted in a better final submission.

\bibliography{/Active/myMaster}

\begin{deluxetable}{l l r l c c c c c c}
\tablecaption{Parallaxes from \HST/FGS astrometry\label{tbl-allP}}
\tablewidth{0in}
\tablehead{
\colhead{\#} &  
\colhead{ID} &
\colhead{$\pi_{abs}$} & &
\colhead{\% error} &
\colhead{ref.\tablenotemark{a}} &
\colhead{$A_V$} &
\colhead{m-M} &
\colhead{$M_K(0)$} &
\colhead{$(V-K)(0)$}
}
\startdata
1&$\ell$ Car&2.01&$\pm$0.20&10.0&{Ben07}&0.52&8.48&-7.55&2.20\\
2&$\zeta$ Gem&2.78&0.18&6.5&{Ben07}&0.06&7.78&-5.73&1.69\\
3&$\beta$ Dor&3.14&0.16&5.1&{Ben07}&0.03&7.52&-5.62&1.59\\
4&$\delta$ Cep\tablenotemark{b}&3.66&0.15&4.1&{Ben07}&0.23&7.18&-4.91&1.44\\
5&FF Aql&2.81&0.18&6.4&{Ben07}&0.64&7.76&-4.39&1.34\\
6&RT Aur&2.40&0.19&7.9&{Ben07}&0.20&8.10&-4.25&1.36\\
7&W Sgr&2.28&0.20&8.8&{Ben07}&0.37&8.21&-5.51&1.54\\
8&X Sgr&3.00&0.18&6.0&{Ben07}&0.58&7.61&-5.15&1.47\\
9&Y Sgr&2.13&0.29&13.6&{Ben07}&0.67&8.36&-5.00&1.58\\
10&T Vul&1.90&0.23&12.1&{Ben07}&0.34&8.61&-4.57&1.3\\
11&$\kappa$ Pav&5.57&0.28&5.0&{Ben11}&0.05&6.27&-3.52&1.65\\
12&VY Pyx&6.44&0.23&3.6&{Ben11}&0.15&5.96&-0.26&1.44\\
13&RZ Cep&2.12&0.16&7.5&{Ben11}&0.78&8.37&-0.4&0.67\\
14&XZ Cyg&1.67&0.17&10.2&{Ben11}&0.30&8.89&-0.29&0.7\\
15&SU Dra&1.42&0.16&11.3&{Ben11}&0.03&9.24&-0.73&1.13\\
16&RR Lyr\tablenotemark{b}&3.77&0.13&3.4&{Ben11}&0.13&7.12&-0.65&1.19\\
17&UV Oct&1.71&0.10&5.8&{Ben11}&0.28&8.84&-0.6&0.95\\
18&HD 213307\tablenotemark{b}&3.65&0.15&4.1&{Ben02b}&0.23&7.19&-0.86&-0.24\\
19&vA 627\tablenotemark{b}&21.74&0.25&1.1&{McA11}&0.00&3.31&3.86&2.38\\
20&vA 310\tablenotemark{b}&20.13&0.17&0.8&{McA11}&0.00&3.48&4.09&2.44\\
21&vA 472\tablenotemark{b}&21.70&0.15&0.7&{McA11}&0.00&3.32&3.69&2.02\\
22&vA 645\tablenotemark{b}&17.46&0.21&1.2&{McA11}&0.00&3.79&4.11&3.11\\
23&vA 548\tablenotemark{b}&20.69&0.17&0.8&{McA11}&0.00&3.42&4.13&2.79\\
24&vA 622\tablenotemark{b}&24.11&0.30&1.2&{McA11}&0.00&3.09&5.13&3.68\\
25&vA 383\tablenotemark{b}&21.53&0.20&0.9&{McA11}&0.00&3.33&5.01&3.85\\
26&Barnard\tablenotemark{b}&545.40&0.30&0.1&{Ben99}&0.00&-3.68&8.21&4.98\\
27&Proxima Cen\tablenotemark{b}&768.70&0.30&0.0&{Ben99}&0.00&-4.43&8.81&6.75\\
28&P3179&7.45&0.16&2.1&{Sod05}&0.14&5.64&3.04&1.39\\
29&P3063&7.43&0.16&2.2&{Sod05}&0.14&5.65&4.70&3.20\\
30&P3030&7.41&0.18&2.4&{Sod05}&0.14&5.65&4.98&3.37\\
31&GJ1005A&166.60&0.30&0.2&{Ben16}&0.00&-1.11&7.80&4.9\\
32&GJ1005B&166.60&0.30&0.2&{Ben16}&0.00&-1.11&9.03&6.09\\
33&GJ22A&99.20&0.60&0.6&{Ben16}&0.00&0.02&6.19&4.29\\
34&GJ22C&99.20&0.60&0.6&{Ben16}&0.00&0.02&8.12&5.44\\
35&GJ1081A&65.20&0.37&0.6&{Ben16}&0.00&0.93&6.78&4.5\\
36&GJ1081B&65.20&0.37&0.6&{Ben16}&0.00&0.93&7.74&5.21\\
37&GJ234A&240.98&0.40&0.2&{Ben16}&0.00&-1.91&7.63&5.37\\
38&GJ234B&240.98&0.40&0.2&{Ben16}&0.00&-1.91&9.21&6.87\\
39&G250-029A&95.59&0.28&0.3&{Ben16}&0.00&0.10&6.60&4.24\\
40&G250-029B&95.59&0.28&0.3&{Ben16}&0.00&0.10&7.63&4.82\\
41&GJ469A&76.41&0.46&0.6&{Ben16}&0.00&0.58&6.73&4.75\\
42&GJ469B&76.41&0.46&0.6&{Ben16}&0.00&0.58&7.74&5.33\\
43&GJ623A\tablenotemark{b}&125.00&0.30&0.2&{Ben16}&0.00&-0.48&6.47&4.3\\
44&GJ623B\tablenotemark{b}&125.00&0.30&0.2&{Ben16}&0.00&-0.48&9.34&6.71\\
45&GJ748A\tablenotemark{b}&98.40&0.30&0.3&{Ben16}&0.00&0.04&6.60&4.49\\
46&GJ748B\tablenotemark{b}&98.40&0.30&0.3&{Ben16}&0.00&0.04&7.69&5.23\\
47&GJ1245A&219.90&0.50&0.2&{Ben16}&0.00&-1.71&8.92&6.15\\
48&GJ1245C&219.90&0.50&0.2&{Ben16}&0.00&-1.71&9.95&8.41\\
49&GJ831A\tablenotemark{b}&125.30&0.30&0.2&{Ben16}&0.00&-0.49&7.18&5.37\\
50&GJ831B\tablenotemark{b}&125.30&0.30&0.2&{Ben16}&0.00&-0.49&8.38&6.27\\
51&G193-027A&110.20&1.10&1.0&{Ben16}&0.00&-0.21&8.67&4.82\\
52&G193-027B&110.20&1.10&1.0&{Ben16}&0.00&-0.21&8.78&5.01\\
53&GJ791.2A&113.40&0.20&0.2&{Ben16}&0.00&-0.27&7.87&5.76\\
54&GJ791.2B&113.40&0.20&0.2&{Ben16}&0.00&-0.27&9.16&7.74\\
55&Feige 24\tablenotemark{b}&14.60&0.40&2.7&{Ben00a}&0.00&4.18&6.38&1.85\\
56&DeHt5&2.9&0.15&5.2&{Ben09}&0.37&7.69&7.84&-0.40\\
57&N7293&4.64&0.27&5.8&{Ben09}&0.09&6.67&7.87&-1.07\\
58&N6853\tablenotemark{b}&2.47&0.16&6.5&{Ben09}&0.30&8.04&2.54&3.00\\
59&A31&1.61&0.21&13.0&{Ben09}&0.10&8.97&6.69&-0.44\\
60&V603 Aql&4.01&0.14&3.5&{Har13}&0.10&6.98&4.37&0.20\\
61&DQ Her&2.59&0.21&8.1&{Har13}&0.30&7.93&5.00&1.96\\
62&RR Pic&1.92&0.18&9.4&{Har13}&0.13&8.58&3.67&0.13\\
63&GK Per&2.10&0.12&5.7&{Har13}&1.00&8.39&1.67&3.03\\
64&HP Lib&5.07&0.33&6.5&{Roe07}&0.34&6.47&7.35&-0.91\\
65&CR Boo&2.97&0.34&11.4&{Roe07}&0.03&7.64&8.59&-1.87\\
66&V803 Cen&2.88&0.24&8.3&{Roe07}&0.31&7.70&6.12&-0.48\\
67&AM CVn&1.65&0.30&18.2&{Roe07}&0.05&8.91&1.15&3.91\\
68&GP Com&13.34&0.33&2.5&{Roe07}&0.02&4.37&10.78&1.05\\
69&SS Aur&5.99&0.33&5.5&{Har04}&0.10&6.11&6.30&-2.20\\
70&SS Cyg&7.30&0.20&2.7&{Har16}&0.12&5.68&3.71&-1.80\\
70A&SS Cyg&8.30&0.41&4.9&{Nel13}&0.12&5.39&4.00&-1.80\\
71&U Gem&9.96&0.37&3.7&{Har04}&0.00&5.01&5.91&3.62\\
72&WZ Sge&22.97&0.15&0.7&{Har04}&0.00&3.19&10.87&0.88\\
73&RU Peg&3.55&0.26&7.3&{Har04}&0.00&7.25&3.23&2.14\\
74&V1223 Sgr&1.96&0.18&9.2&{Beu04}&0.47&8.54&4.14&0.08\\
75&EX Hya&15.5&0.29&1.9&{Beu03}&0.00&4.05&7.70&1.51\\
76&TV Col\tablenotemark{b}&2.70&0.11&4.1&{McA01}&0.11&7.84&4.96&0.70\\
77&RW Tri\tablenotemark{b}&2.93&0.33&11.3&{McA99}&0.31&7.67&3.92&0.63\\
78&G166-037&5.2&0.7&13.5&{McC97}&0&6.42&4.36&1.86\\
79&G16-025&3.8&1.0&26.3&{McC97}&0&7.10&4.53&1.70\\
80&HIP 46120&15.01&0.12&0.8&{Cha16}&0.00&4.12&4.34&1.66\\
81&HIP 54639&11.12&0.11&1.0&{Cha16}&0.00&4.77&4.58&2.03\\
82&HIP 87062&8.21&0.11&1.3&{Cha16}&0.19&5.43&3.32&1.63\\
83&HIP 87788&10.83&0.13&1.2&{Cha16}&0.00&4.83&4.63&1.84\\
84&HIP 98492&3.49&0.14&4.0&{Cha16}&0.34&7.29&2.40&1.55\\
85&HIP 103269&14.12&0.10&0.7&{Cha16}&0.00&4.25&4.36&1.66\\
86&HIP 106924&14.47&0.10&0.7&{Cha16}&0.00&4.20&4.37&1.79\\
87&HIP 108200&12.4&0.09&0.7&{Cha16}&0.06&4.53&4.56&1.83\\
88&GJ 876&214.6&0.20&0.1&{Ben02c}&0&-1.66&6.67&5.18\\
89&$\upsilon$ AND&73.71&0.10&0.1&{McA10}&0.00&0.66&2.20&1.24\\
90&HD 38529&25.11&0.19&0.8&{Ben10}&0.00&3.00&1.22&1.68\\
91&$\gamma$ Cep&74.27&0.12&0.2&*&0.00&0.65&0.37&2.20\\
92&HD 47536&8.71&0.16&1.8&*&0.01&5.30&-2.93&2.87\\
93&$\rho^1$ Cnc&79.78&0.30&0.4&{McA04}&0.00&0.49&3.49&1.97\\
94&HD 128311&60.53&0.15&0.2&{McA14}&0.00&1.09&3.99&2.43\\
95&$\epsilon$ Eri&311.37&0.11&0.0&{Ben06}&0.00&-2.47&4.14&2.06\\
96&HD 136118&19.12&0.22&1.2&{Mar09}&0.00&3.59&2.00&1.35\\
97&HD 33636&35.60&0.20&0.6&{Bea07}&0.00&2.24&3.32&1.50\\
98&XO-3&5.67&0.14&2.4&{Joh16}&0.00&6.23&2.56&1.06\\
99&HD 202206&22.98&0.13&0.6&*&0.00&3.19&3.29&1.60\tablenotemark{c}\\
100&HD 84937&12.24&0.20&1.6&{VdB14}&0.02&4.58&2.48&1.26\\
101&HD 132475&10.18&0.21&2.1&{VdB14}&0.02&4.98&1.96&1.61\\
102&HD 140283&17.18&0.26&1.5&{VdB14}&0.01&3.84&1.75&1.62\\
103&HD 140283&17.15&0.14&0.8&{Bon13}&0&3.83&1.76&1.62\\
104&REJ 0317-853&34.38&0.26&0.8&{K\"{u}b10}&0&2.39&13.12&-0.79\\
105&LB 9802&33.28&0.24&0.7&{K\"{u}b10}&&&&\\
\enddata
\tablenotetext{a}{McA10; \cite{McA10}: Ben10; \cite{Ben10}: Ben02b; \cite{Ben02b}: Ben02c; \cite{Ben02c}:McA04; \cite{McA04}: McA14; \cite{McA14}: Ben06; \cite{Ben06}: Mar10; \cite{Mar10}: Bea07; \cite{Bea07}: McA11; \cite{McA11}: Ben99; \cite{Ben99}: Sod05; \cite{Sod05}: Ben16; \cite{Ben16}: Ben00; \cite{Ben00a}: McA99; \cite{McA99}: McA01; \cite{McA01}: Ben09; \cite{Ben09}: Har13; \cite{Har13}: Roe07; \cite{Roe07}: Ben07; \cite{Ben07}: Ben11; \cite{Ben11}: Har04; \cite{Har04}: Har16; \cite{Har16}: Beu03; \cite{Beu03}: Beu04; \cite{Beu04}: McC97; \cite{McC97}: VdB14; \cite{VdB14}: Bon13; \cite{Bon13}; Nel13; \cite{Nel13}; K\"{u}b10; \cite{Kub10}: Joh16; Johns-Krull et al., in preparation: Cha16; Chaboyer et al 2016, submitted: * ; Benedict et al., in preparation
}
\tablenotetext{b}{Original STAT targets.}
\tablenotetext{c}{$(V-K)(0)$ estimated from $SIMBAD$ spectral type.}
\end{deluxetable}

\begin{deluxetable}{c l l c c c c c c}
\tablecaption{\HST/FGS, \G~DR1, and \HIP\tablenotemark{a} Parallaxes\label{tbl-HG}}
\tablewidth{0in}
\tablehead{
\colhead{\#} &  
\colhead{ID} &
\colhead{$V$}&
\colhead{\HSTs $\pi_{abs}$} & 
\colhead{error} &
\colhead{\G~$\pi_{abs}$} &
\colhead{error\tablenotemark{b}} &
\colhead{\HIP~$\pi_{abs}$} &
\colhead{error}
}
\startdata
5&FF Aql&5.36&2.81&0.18&1.64&0.89&2.11&0.33\\
12&VY Pyx&7.25&6.44&0.23&3.85&0.28&5.01&0.44\\
13&RZ Cep&9.53&2.12&0.16&2.65&0.24&0.59&1.48\\
14&XZ Cyg&9.69&1.67&0.17&1.56&0.23&2.29&0.84\\
15&SU Dra&9.84&1.42&0.16&1.43&0.28&0.2&1.13\\
16&RR Lyr&7.88&3.77&0.13&3.64&0.23&3.46&0.64\\
17&UV Oct&9.45&1.71&0.10&2.02&0.23&2.44&0.81\\
20&vA 310&10.04&20.13&0.17&21.76&0.41&19.31&1.93\\
21&vA 472&9.05&21.70&0.15&20.88&0.24&21.86&1.72\\
22&vA 645&11.27&17.46&0.21&22.02&0.25&19.12&5.45\\
25&vA 383&12.20&21.53&0.20&20.99&0.36&&\\
28&P3179&10.05&7.45&0.16&8.29&0.88&&\\
55&Feige 24&12.65&14.60&0.40&13.06&0.76&10.9&3.94\\
60&V603 Aql&11.76&4.01&0.14&2.92&0.54&4.96&2.45\\
62&RR Pic&12.11&1.92&0.18&2.45&0.44&-4.63&1.94\\
70&SS Cyg&11.64&7.30&0.20&8.56&0.33&&\\
70A&SS Cyg&11.64&8.30&0.41&8.56&0.33&&\\
82&HIP 87062&10.67&8.94&0.11&8.38&0.86&9.59&2.21\\
83&HIP 87788&11.72&10.73&0.13&10.97&0.26&10.01&2.79\\
84&HIP 98492&11.47&3.45&0.14&2.48&0.37&9.78&2.77\\
85&HIP 103269&10.33&14.12&0.10&13.76&0.22&14.86&1.31\\
92&HD 47536&5.26&8.71&0.16&7.95&0.57&8.11&0.23\\
96&HD 136118&6.95&19.12&0.22&19.22&0.27&21.47&0.54\\
97&HD 33636&6.99&35.60&0.20&33.96&0.82&35.25&1.02\\
98&XO-3&9.86&5.67&0.14&4.96&0.35&&\\
99&HD 202206&8.08&22.98&0.13&21.94&0.26&22.06&0.82\\
101&HD 132475&8.55&10.18&0.21&10.62&0.24&10.23&0.84\\
102&HD 140283&7.21&17.18&0.26&15.96&0.43&17.16&0.68\\
\hline
mean error&&&&0.18&&0.41&&1.58\\
median error&&&&0.17&&0.33&&1.31\\
\enddata
\tablenotetext{a}{From \cite{Lee07a}.}
\tablenotetext{b}{Does not include estimated systematic error of 0.3 mas as stated on the \G~DR1 website.}
\end{deluxetable}

\begin{landscape}
\begin{deluxetable}{l l r r c c c r r c c}
\tablecaption{Companion Masses from \HST-FGS astrometry\label{tbl-allM}}
\tablewidth{0in}
\tablehead{
\colhead{Companion} &  
\colhead{$\cal{M}_\sun$} &
\colhead{[Fe/H]} & 
\colhead{Sp. T.} &
\colhead{d [pc]} &
\colhead{ecc} &
\colhead{$\cal{M}$ [Jup]} &
\colhead{$\alpha$ [mas]} &
\colhead{inc [\arcdeg]} &
\colhead{P [d]} &
\colhead{ref\tablenotemark{a}}
}
\startdata
GJ 876 b&0.32&-0.12 2&M4 V&4.7&0.1&1.89$\pm$0.34&0.25$\pm$0.06&84$\pm$6&61&B02\\
55 Cnc d&1.21&+0.32&G8 V&12.5&0.33&4.9~1.1&1.9&53~7&4517&M4\\
$\epsilon$ Eri b&0.83&-0.03&K2 V&3.2&0.7&1.6~0.2&1.9 0.2&30~4&2502&B06\\
HD 33636 B&1.02&-0.13&G0 V&28.1&0.48&142~11 &14.2 0.2&4.0~0.1&2117&B07\\
$\upsilon$ And c&1.31&+0.15&F8 V&13.5&0.25&14~4&0.62 0.08&8~1&241.2&M10\\
$\upsilon$ And d&|&|&|&|&0.32&10 ~ 2&1.39 0.07&24~1&1281.5&\\
HD 136118 b&1.24&-0.01&F9 V&52.3&0.35&42~15&1.45 0.25&163.1~3&1191&Ma10\\
HD 38529 c&1.48&+0.27&G4 IV&40.0&0.36&17.6~1.4&1.05 0.06&48.3~3.7&2136&B10\\
HD 128311 c&0.83&0.2&K0 V&16.5&0.154&3.79~0.7&0.46 0.09&56~14&921.5&M14\\
\enddata
\tablenotetext{a}{M10; \cite{McA10}:B10; \cite{Ben10} :B02; \cite{Ben02b}:M4; \cite{McA04}:M14; \cite{McA14}:B6; \cite{Ben06}:Ma10; \cite{Mar10}:B07; \cite{Bea07}}
\end{deluxetable}
\end{landscape}

\begin{figure}
\includegraphics[width=6in]{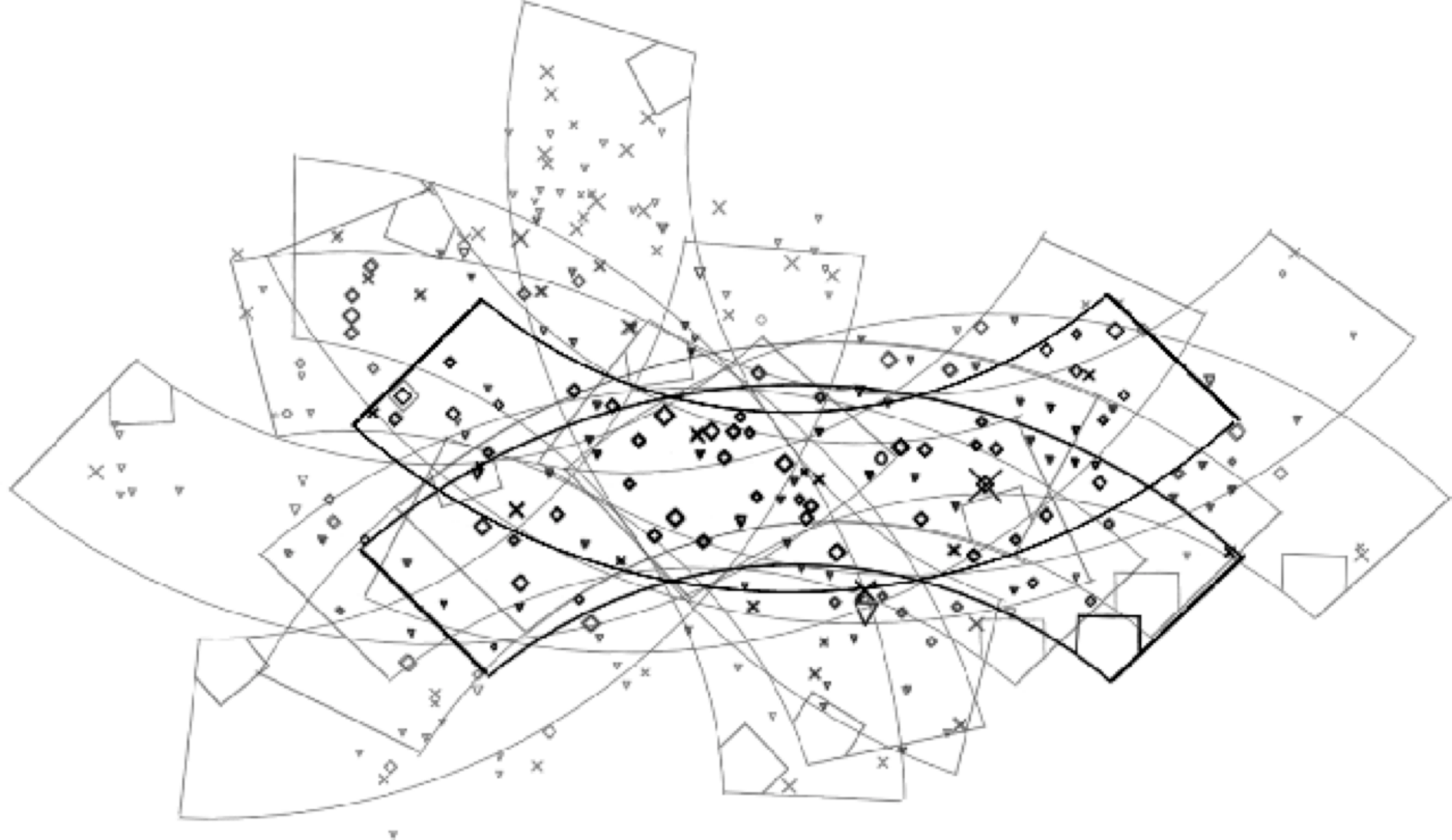}
\caption{Positioning of the FGS\,1r pickle on the M35 calibration field, showing offsets and various roll angles. M35 member stars observed are shown with symbol size proportional to V magnitude. Stars actually observed for the OFAD have \textbf {bold} symbols. The entire observation sequence is completed in about two days with no hiatus. See \cite{McA06} for details.\label{Fig-OFAD}}
\end{figure}

\begin{figure}
\includegraphics[width=5.5in]{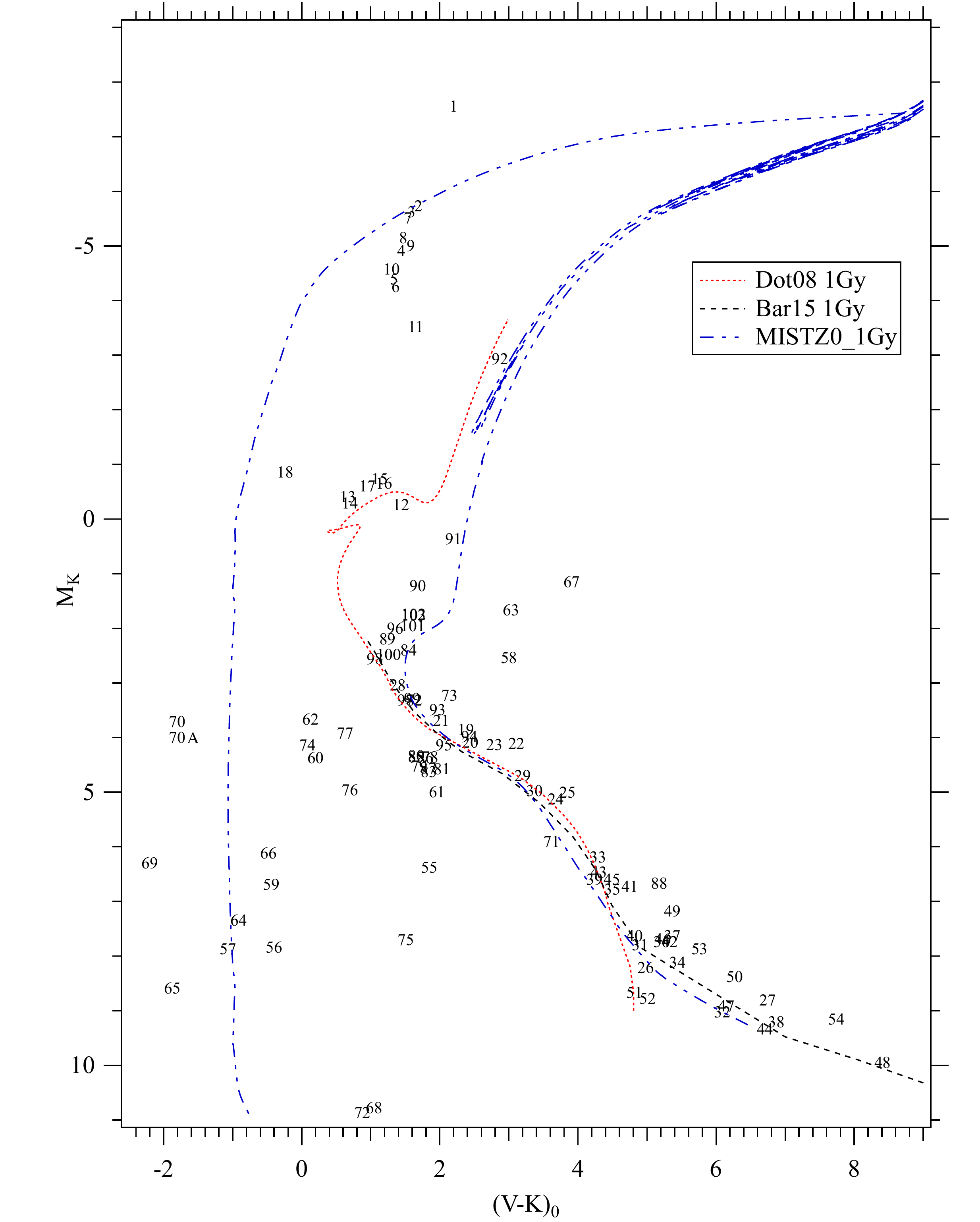}
\caption{An $M_K$ vs. $(V-K)_0$  Hertzsprung-Russell diagram constructed from the  parallaxes in Table~\ref{tbl-allP}, plotted with ID \textbf{numbers}. Included are Cepheids and RRLs; main sequence stars; post-Main Sequence sub-giants, WDs, CVs, novae; and Pop II main sequence stars. Also plotted are several recent stellar model isochrons; Dot08 = \citep{Dot08}; BAR15 = \citep{Bar15}, MIST = \citep{Dot16, Cho16}. All have solar metallicity and map a stellar age of 1Gy. Given the parallax precision and small effect of absorption, all the scatter along the Main Sequence is due to cosmic dispersion, intrinsic to the stars. \label{Fig-HR}}
\end{figure}

\begin{figure}
\includegraphics[width=5in]{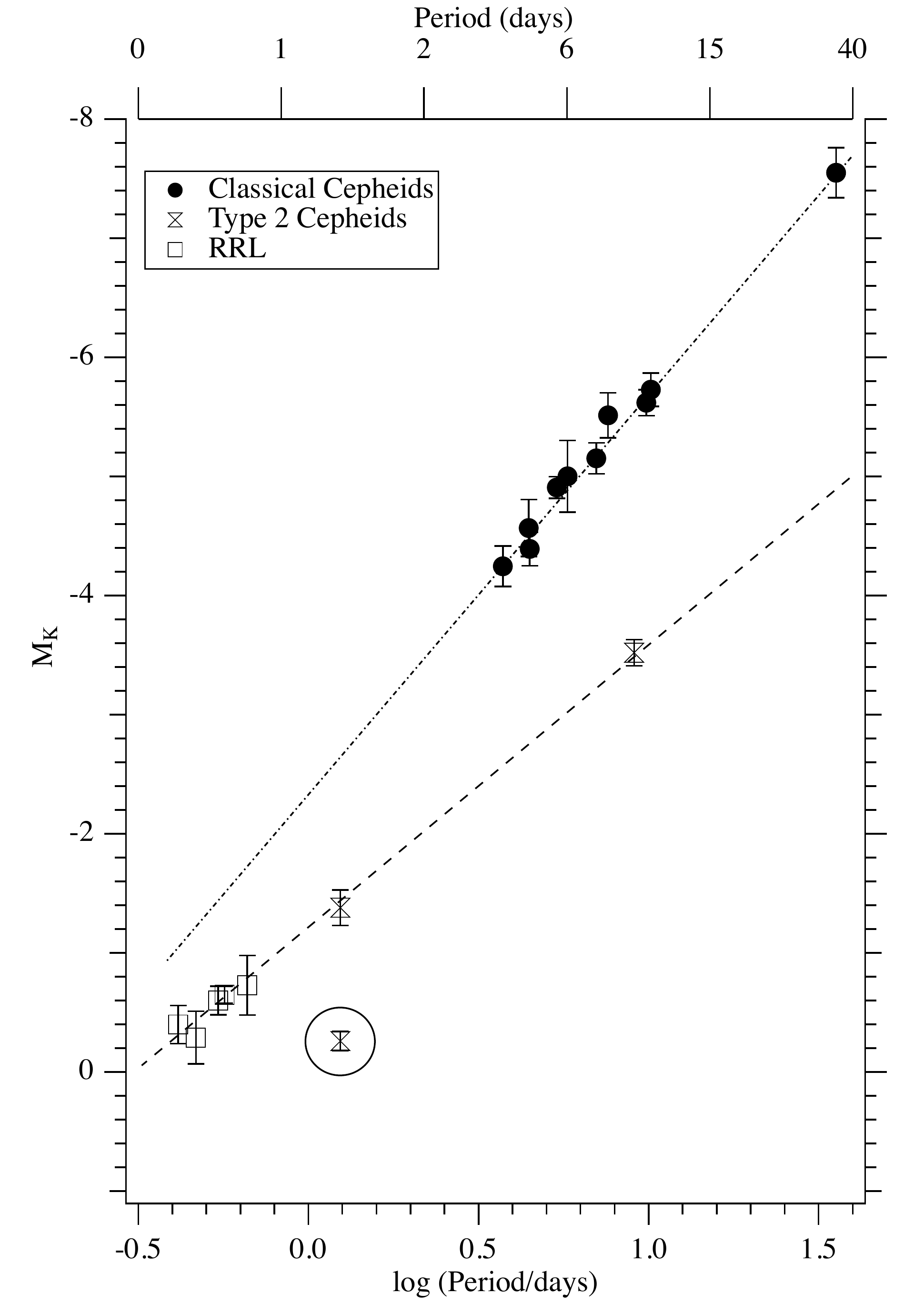}
\caption{Period-Luminosity relations constructed with absolute magnitudes, $M_K$, derived with \HST /FGS parallaxes. Classical Cepheids are from \cite{Ben07}, and RR Lyr and dwarf Cepheid stars from \cite{Ben11}. The type 2 Cepheid $\kappa$ Pav agrees with a previously determined PLR slope \citep{Sol06}. The \HSTs VY Pyx (circled) value does not, but the \G~parallax yields $M_K$ (not circled at logP=0.09) in agreement.  \label{Fig-PLR}}
\end{figure}

\begin{figure}
\includegraphics[width=5in]{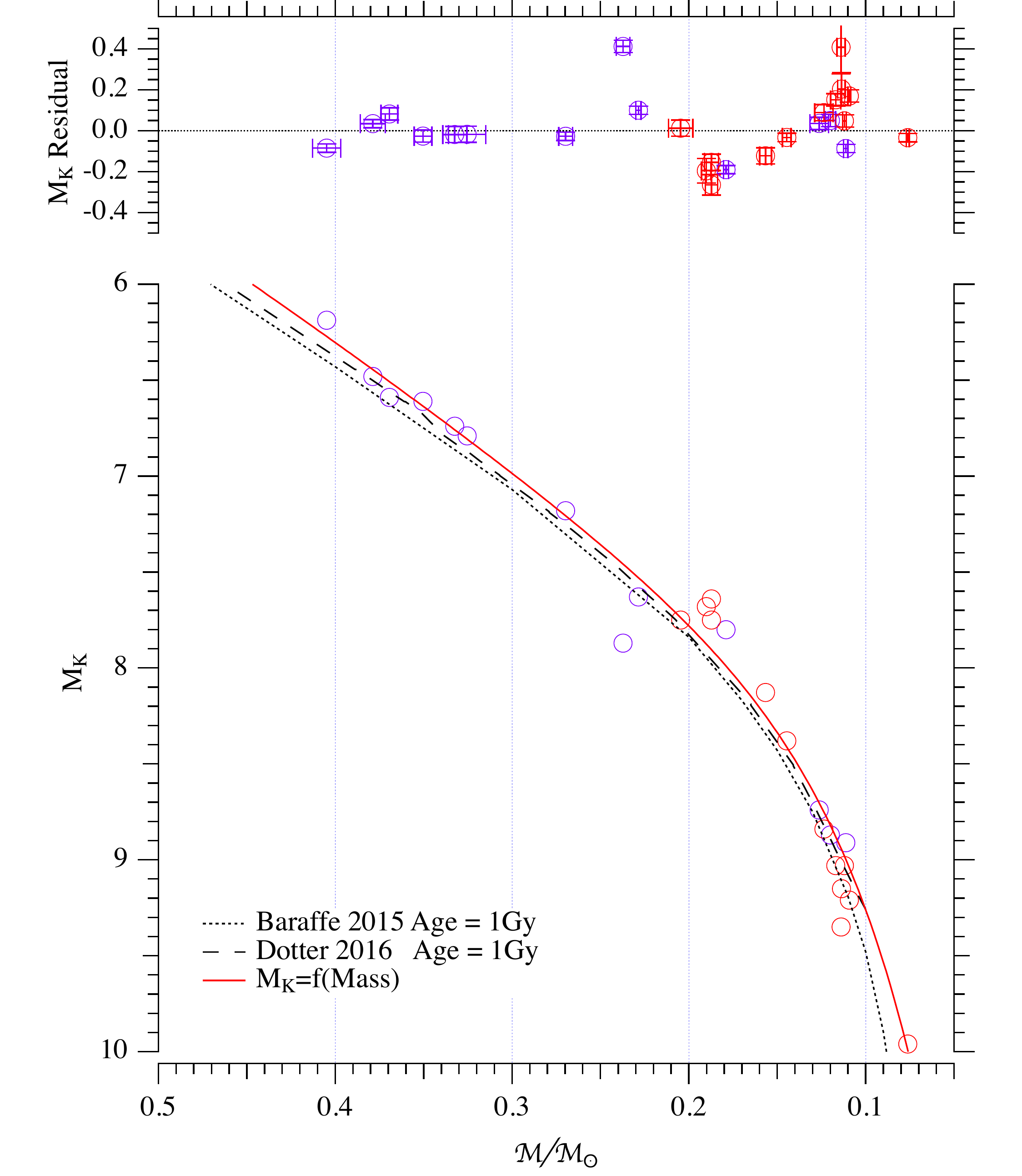}
\caption{A $K$-band Mass-Luminosity relation for low-mass stars constructed using only  component masses  determined by a combination  of POS and TRANS mode \HST /FGS astrometry. Blue symbols are primary components; red symbols secondary components. The plot contains a subset of the stars comprising the MLR in \cite{Ben16}. The fitting function, f(Mass), is the double exponential with offset described in \cite{Ben16}. Also plotted are solar-metallicity models from \cite{Dot16} and \cite{Bar15}.\label{Fig-MLR}}
\end{figure}

\begin{figure}
\includegraphics[width=5in]{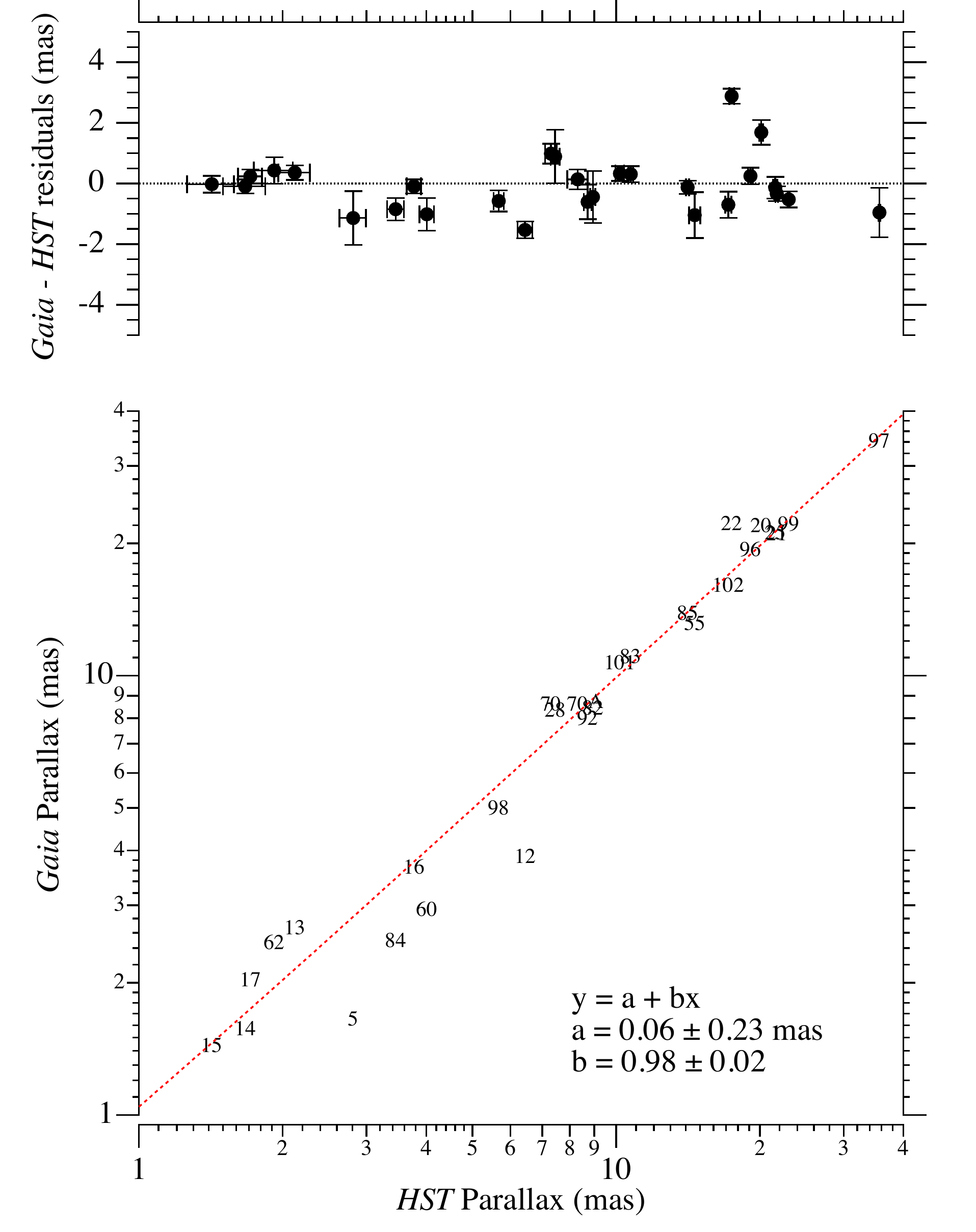}
\caption{\G~ parallaxes from \G~DR1 compared to parallaxes from \HST /FGS (Table~\ref{tbl-HG}). The regression line is impartial, in that errors in both \G~ and \HSTs parallaxes are considered (Jefferys et al. 1988).  The residuals are tagged with \HSTs errors on the x axis and \G~errors on the y axis. Notably large residuals  include FF Aql (\textbf{5}),VY Pyx (\textbf{12}), VA 645 (\textbf{22}), vA 310 (\textbf{20}), and HD 33636 (\textbf{97}).}\label{Fig-GH}
\end{figure}

\begin{figure}
\includegraphics[width=6in]{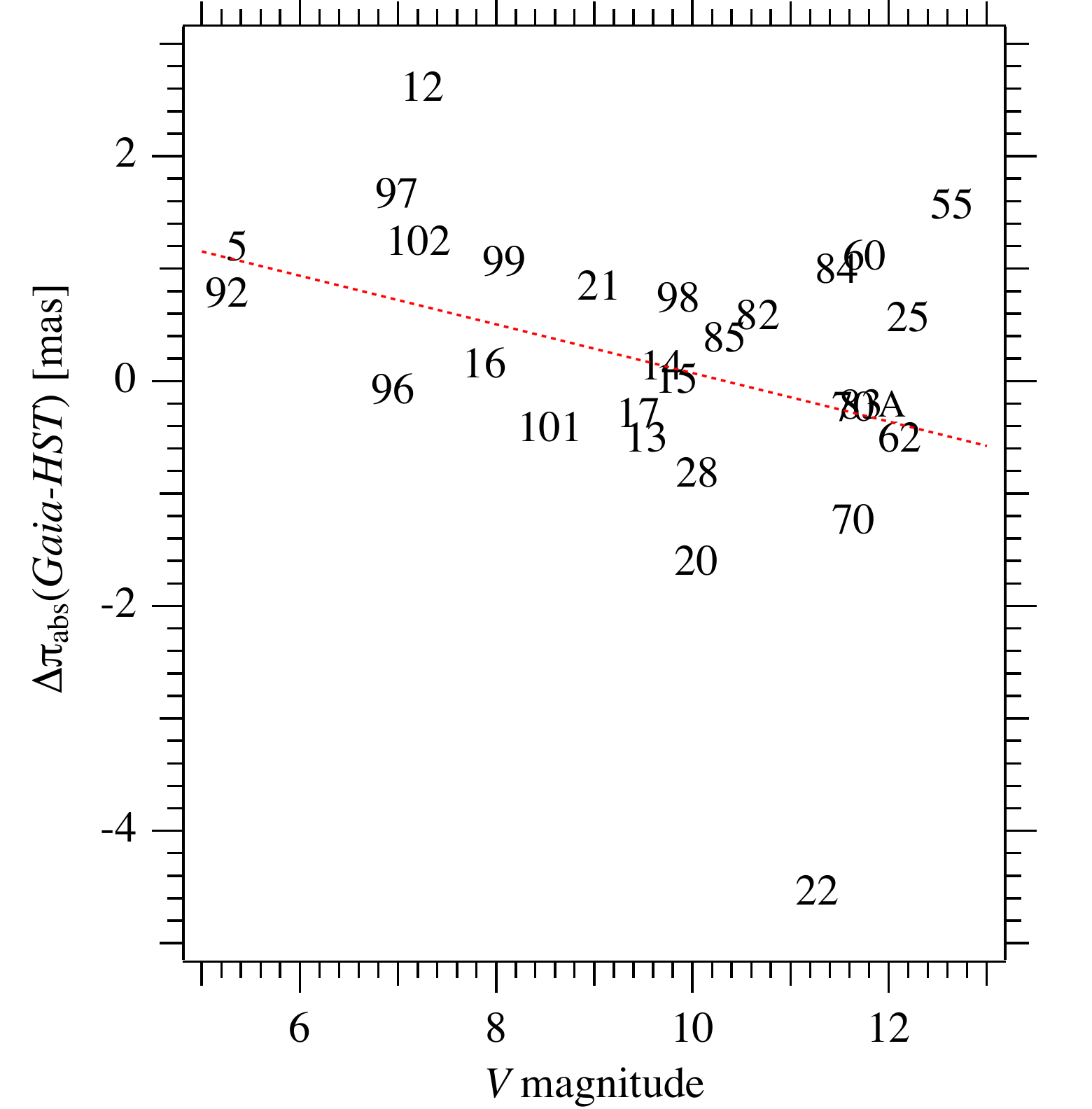}
\caption{\G~and \HSTs parallax  (from Table~\ref{tbl-HG}) differences as a function of magnitude. The regression line has Pearson's r = -0.34, indicating a weak correlation; in general larger differences for brighter stars. The significant outlier at $V=11.2$ is the Hyad vA 645 (\textbf{22}).}\label{fig-GHV}
\end{figure}

\end{document}